\documentclass[%
 reprint,
 amsmath,amssymb,
 aps,
]{revtex4-2}

\usepackage{graphicx}
\usepackage{dcolumn}
\usepackage{bm}
\usepackage{color}

\begin{document}

\preprint{APS/123-QED}

\title{Membrane buckling and the determination of Gaussian curvature modulus}
\author{Mei-Ting Wang}
\author{Rui Ma}
 \email{ruima@xmu.edu.cn}
\altaffiliation{
Fujian Provincial Key Lab for Soft Functional Materials Research,
Research Institute for Biomimetics and Soft Matter,
Department of Physics, College of Physical Science and Technology,
Xiamen University, Xiamen 361005, People’s Republic of China
}%

\author{Chen-Xu Wu}
 \email{cxwu@xmu.edu.cn}
\affiliation{%
 Fujian Provincial Key Lab for Soft Functional Materials Research,
Research Institute for Biomimetics and Soft Matter,
Department of Physics, College of Physical Science and Technology,
Xiamen University, Xiamen 361005, People’s Republic of China
}



\date{\today}

\begin{abstract}
Biological membranes are able to exhibit various morphology due to the fluidity of the lipid molecules within the monolayers. The shape transformation of membranes has been well described by the classical Helfrich theory, which consists only a few phenomenological parameters, including the mean and the Gaussian curvature modulus. Though various methods have been proposed to measure the mean curvature modulus, determination of the Gaussian curvature modulus remains difficult both in experiments and in simulation. In this paper we study the buckling process of a rectangular membrane and a circular membrane subject to compressive stresses and under different boundary conditions. We find that the buckling of a rectangular membrane takes place continuously, while the buckling of a circular membrane can be discontinous depending on the boundary conditions. Furthermore, our results show that the stress-strain relationship of a buckled circular membrane can be used to effectively determine the Gaussian curvature modulus.
\end{abstract}

\maketitle


\section{introduction}
Buckling, a common phenomenon observed in our daily life~\cite{brush1975buckling,cerda2003geometry,2006Soft,khang2009mechanical,yin2008stress,domokos2003symmetry,vliegenthart2006forced,lobkovsky1995scaling,moldovan1999buckling,jain2021compression,zhang2022general,ebrahimi2022buckling}, refers to the sudden change in shape of an elastic object under compressive loads. The research of buckling can be dated back to as early as 1691, when Jacob Bernoulli studied the buckling of an elastic beam~\cite{timovsenko1961theory}. In the 18th century, Leonhard Euler and Daniel Bernoulli further developed the elastic beam theory which nowadays constitutes an important branch of the continuum mechanics, and has broad applications in structural and mechanical engineering~\cite{1953History}. 

In recent years, buckling in fluid membranes has drew attention of many physicists~\cite{noguchi2011anisotropic,hu2013determining,stecki2004correlations,den2005area,noguchi2006meshless,2020On,2021Elastic,pinigin2022determination,lac2004spherical,lenz2009membrane}. The major components of a fluid membrane are lipid molecules, which are typically composed of a hydrophilic polar head and two hydrophobic hydrocarbon tails. In aqueous solution, the lipid molecules assemble into a double layer structure with the hydrophobic tails embedded inside so as to avoid water and the hydrophilic heads exposed to the water, which is referred to as a lipid bilayer~\cite{1989Biomembranes}. The lipid molecules can freely move laterally within the monolayer, which differs the fluid membrane from the solid shell. The buckling of a lipid membrane therefore exhibits novel behaviors compared with that of the elastic shell, such as the anisotropic tension and negative compressibility~\cite{noguchi2011anisotropic,hu2013determining}. 

The thickness ($\sim 4\mathrm{nm}$) of a typical fluid membrane is negligible compared with its lateral dimension. When considering the deformation of the membrane on length scales that are even moderately larger than the thickness, the Helfrich theory~\cite{Helfrich1973,helfrich1974size}, which treats the membrane as a continuum 2D surface, has been extremely successful in many applications~\cite{evans1990entropy,baumgart2011thermodynamics,julicher1993domain,seifert1997configurations,lipowsky2013spontaneous,seifert1991shape,julicher1996shape}. The theory only has a couple of phenomenological parameters to characterize the membrane's property, such as the mean and the Gaussian curvature modulus $\kappa$ and $\bar{\kappa}$, and the spontaneous curvature $c_0$. The local deformation of a membrane directly depends on the mean curvature modulus $\kappa$, which makes the measurement of $\kappa$ a relatively easy assay. In experiments, $\kappa$ is typically obtained via the fluctuation spectrum of a flat membrane~\cite{brochard1975frequency,schneider1984thermal,schneide1984thermal}, or the force spectrum to pull a membrane tether from a spherical shape~\cite{faucon1989bending,henriksen2004vesicle}. However, measurement the Gaussian curvature modulus $\bar{\kappa}$ is difficult because it only depends on the global deformation of the membrane that alters the topology due to the Gauss-Bonnet theorem~\cite{kreyszig2013differential,do2016differential}. For this very reason, in cellular processes which change the topology of the membrane, such as cell division, endocytosis and exocytosis, the role of Gaussian curvature modulus $\bar{\kappa}$ can not be simply ignored~\cite{lodish2008molecular,mcnew2013gtp,lipowsky2022remodeling}. 

Boundary conditions (BCs) are important to determine the shape of an elastic object. For a fluid membrane patch with an open edge, the Gaussian curvature modulus $\bar{\kappa}$ has a contribution to the boundary stress. This dependence has been used by Molecular dynamics (MD) simulations to estimate the Gaussian curvature modulus $\bar{\kappa}$ either via the closure of a spherical vesicle~\cite{hu2013gaussian} or via the edge fluctuations of a flat membrane ~\cite{zelisko2017determining}. However, both methods require the simulation of a large membrane either for a long time or with multiple repeats, and therefore are computational expensive.

Recently, buckling of a flat membrane has been suggested as a MD protocol to measure the mean curvature modulus $\kappa$ of the membrane~\cite{hu2013determining}. Its buckling protocol is easy to implement, and has been shown to be robust against the coarse graining level of the lipid models and the treatment of the solvent when estimating $\kappa$. Inspired by the work, in this paper, we consider two kinds of geometries of membrane, i.e. a rectangular membrane and a circular one, and respectively investigate their buckling phenomena under a compressive stress via the Helfrich theory~\cite{Helfrich1973,helfrich1974size}. Our numerical results of the stress-strain relationship for the rectangular membrane agree well with the theoretical predictions reported in the literature~\cite{hu2013determining}. Moreover, through investigating the circular membrane, we find that the buckling process shows qualitatively different behaviors under distinct BCs. The stress-strain relationship under the free-hinge BC strongly depends on the Gaussian curvauture modulus $\bar{\kappa}$, which therefore provides an effective method to determine the Gaussian bending rigidity of a fluid membrane through a buckling protocol.           

\section{Theoretical model}
We model the rectangular and the circular membranes as a one-dimensional and an axis-symmetric two-dimensional patches, respectively. We consider two types of BCs, namely the free-hinge BC in which the membrane is allowed to rotate freely at the edge where compressible stresses are applied, and the fixed-hinge BC in which the membrane angle is fixed to be in parallel with the substrate at the edge.   
\subsection{The rectangular membrane}
\begin{figure}[h]
\centering
\includegraphics[scale=0.46]{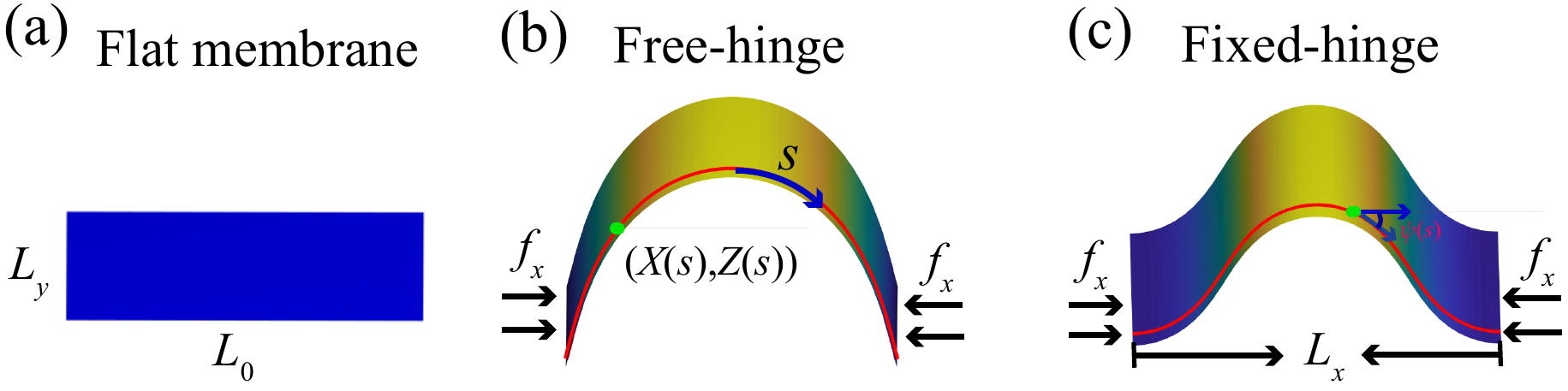}
\caption{Illustration of a rectangular membrane deformed from (a) a flat shape, to a buckled shape under (b) free-hinge BC or (c) fixed-hinge BC.}
\label{rectmembrane}
\end{figure}

For the rectangular membrane which is initially laid flat at the horizontal surface with a length $L_0$ in the $x$-direction and a width of $L_y$ in the $y$-direction, the buckling is driven by a compressive stress $f_x$ applied at the two ends of the membrane along the $x$-direction. When the stress $f_x$ exceeds a critical value, the membrane will buckle and the shape is depicted by the coordinates $[X(s),Z(s)]$, where $s$ is the arclength along the $x$-direction (Fig.~\ref{rectmembrane}). We assume the total arclength $L_{0}$ of the membrane is invariant during buckling process and introduce $\psi(s)$ as the angle made between the tangential direction of the arc and the $x$-direction. The total energy of the membrane is written as
\begin{equation}
E=\frac{\kappa}{2}\int{ c_1 ^2}\text{d}A+f_{x}L_{x}L_{y},
\label{recttotalenergy}
\end{equation}
where the first term is the bending energy, with $c_1 = \text{d}\psi/\text{d}s$ the principal curvature along the arc direction, and $\text{d}A = L_y \text{d}s$ the area element. The second term is the boundary energy, where the compressive stress $f_x$ is essentially a Lagrangian multiplier with a unit of force per unit length imposed on the base length $L_x$ in the $x$-direction. Note that the Gaussian curvature for the curved rectangular membrane is zero, and therefore has no contribution to the total energy. 

\subsection{The  circular membrane}
For the circular membrane which is initially laid flat with a radius of $R_0$, the buckling is driven by a compressive stress $f_r$ along the radial direction applied at the perimeter of the membrane. When the stress exceeds a critical value, the membrane will buckle with the shape of the membrane depicted by the coordinates $[R(s),Z(s)]$, where $s$ is the arclength along the meridian direction, as shown in Fig.~\ref{fig2}. Different from the rectangular membrane, the total arclength $S$ can change during buckling, but the total area $A$ of the membrane remains invariant, the same as in the rectangular membrane. The angle $\psi(s)$ spans between the tangential meridian direction and the radial direction. The total energy of the membrane reads
\begin{equation}
F=\int\left[\frac{\kappa}{2}\left( c_1+c_2 \right) ^2+\bar{\kappa}c_1c_2\right]\text{d}A+\sigma A+f_{r} \pi R_{b}^2,
\label{cirtotalenergy}
\end{equation}
where the first integral is the bending energy with $c_1 = \text{d}\psi/\text{d}s$ and $c_2 = \sin\psi/R$ the two principal curvatures of the membrane surface, and the two terms in the square brackets represent contributions from the mean curvature and the Gaussian curvature, respectively. The Lagrangian multiplier $\sigma$ in the second term is to impose a constant surface area condition for the membrane during the buckling process, which can be interpreted as the membrane tension. The last term is the boundary energy, where the compressive stress $f_{r}$ is also a Lagrangian multiplier imposed on the base area $\pi R_{b}^2$ with $R_{b}$ the base radius of the membrane when it buckles.

\subsection{Boundary conditions}
\begin{figure}[h]
\centering
\includegraphics[scale=0.26]{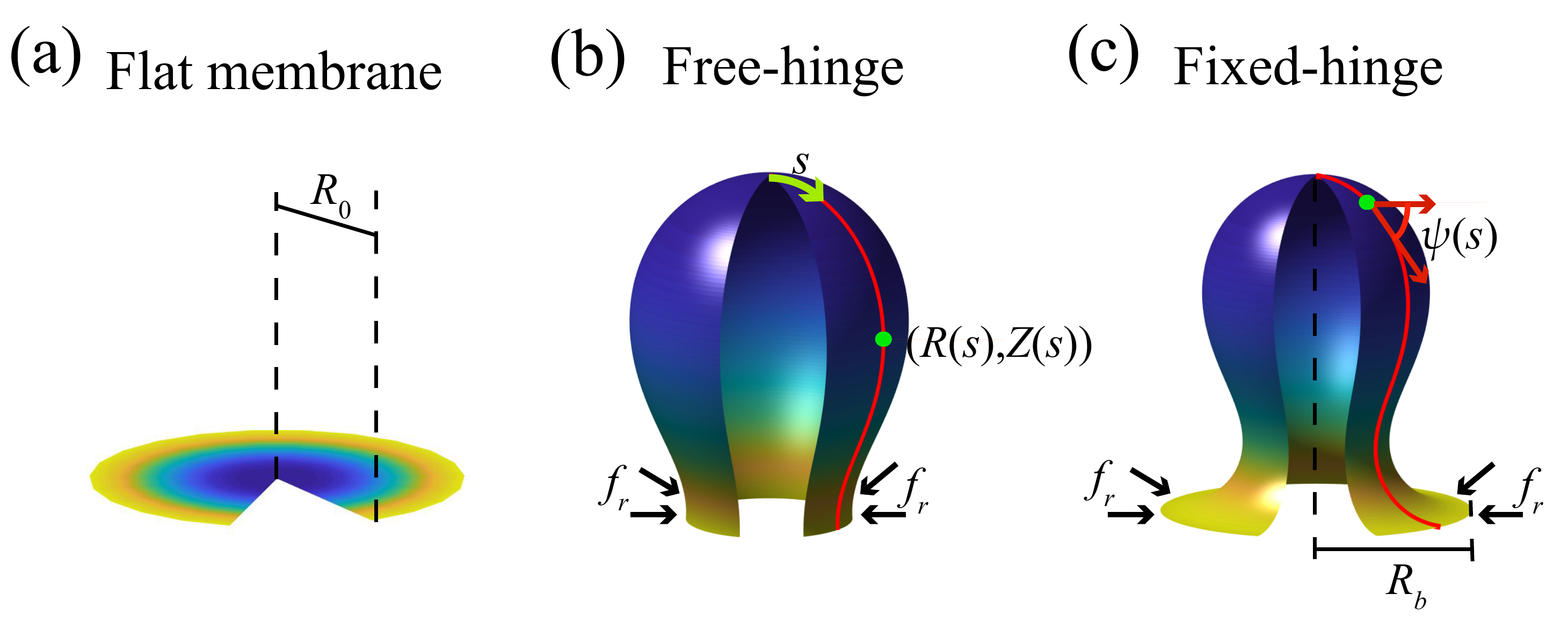}
\caption{Illustration of a circular membrane deformed from (a) a flat circular shape, to a buckled shape under (b) free-hinge BC or (c) fixed-hinge BC.}
\label{fig2}
\end{figure}

For the free-hinge BC, we have the vanishing bending moment at the boundary. In the case of the rectangular membrane, it is expressed as the vanishing curvature $c_1 = 0$. In the case of the circular membrane (Fig.~\ref{fig2}(b)), it implies that
\begin{equation}
\label{eq:freehinge}
    \kappa(c_1+c_2)+\bar{\kappa}c_2 = 0.
\end{equation}
Note that the Gaussian curvature modulus $\bar{\kappa}$ appears in Eq.~(\ref{eq:freehinge}). This is the key reason why we can use the buckling protocol to measure $\bar{\kappa}$ for the free-hinge BC, which will be elaborated later. 

For the fixed-hinge BC, we simply fix the membrane angle $\psi = 0$ unless otherwise stated, as illustrated in Fig.~\ref{fig2}(c) in the case of the circular membrane.    

\section{Results and discussion}
\subsection{The buckling of a rectangular membrane}
The buckling of a rectangular membrane with a fixed-hinge BC has been studied in Ref.~\cite{hu2013determining, noguchi2011anisotropic} both analytically and via MD simulations. Here we numerically solve the shape equations and analyze the buckling process from an energetic point of view and incorporate the free-hinge BC into consideration. 

\begin{figure}[!ht]
\centering
      \includegraphics[scale=0.45]{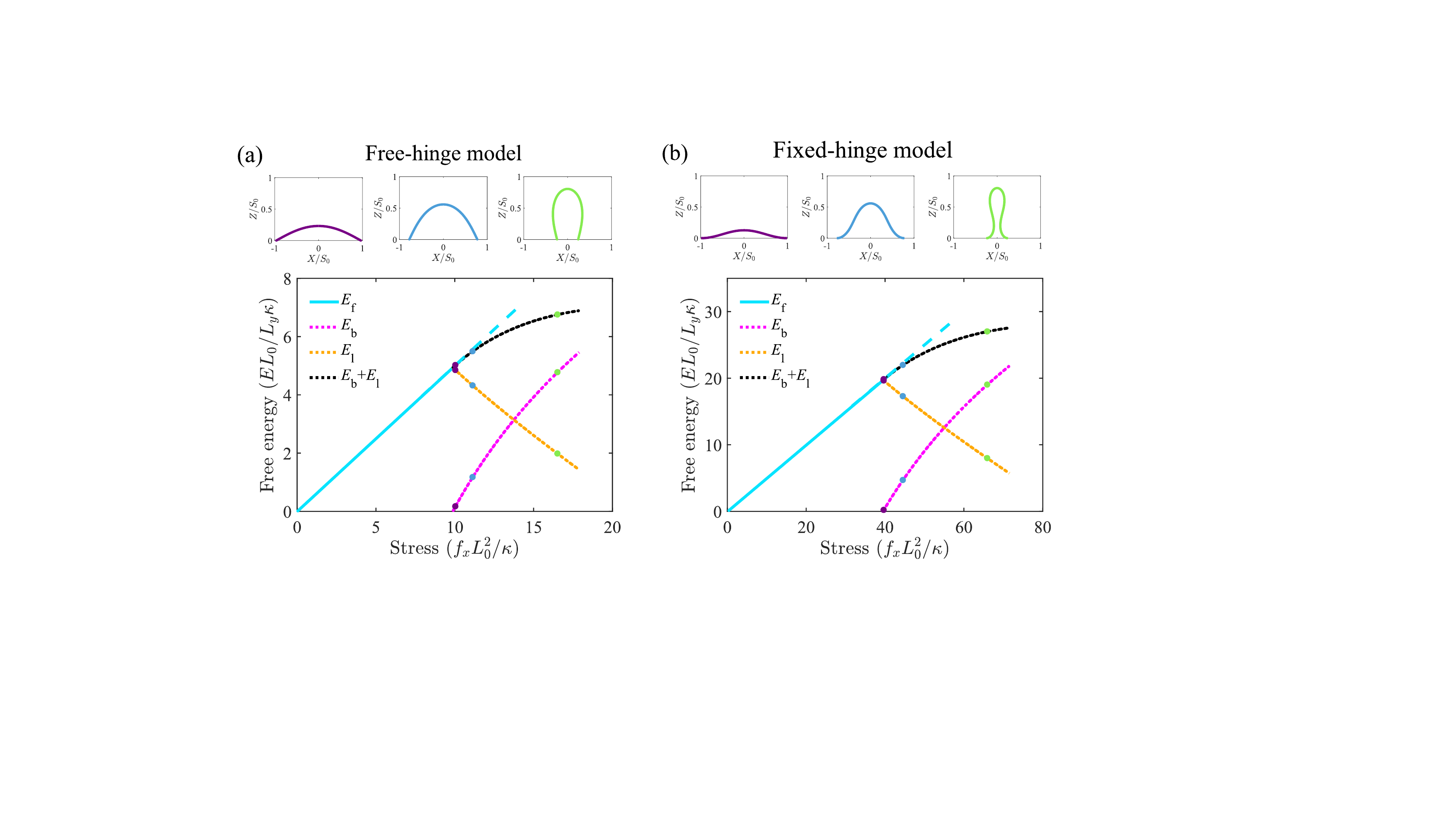}
\caption{Effect of the stress on the free energy of the rectangular membrane based on (a) a free-hinge or (b) a fixed-hinge model. Here, $S_{0}=L_{0}/2$.}
\label{fig3}
\end{figure}

The flat shape is always a trivial solution to the shape equations regardless of the stress $f_x$ and the BCs, and the total energy $E_{\text{f}}$ of the flat shape increases with the stress $f_x$ linearly due to the boundary energy (cyan line in Fig.~\ref{fig3}). When the stress $f_x$ exceeds a critical value, a new branch of solutions emerge due to membrane buckling (Fig.~\ref{fig3}, black dotted line). The total energy of the buckled shape is lower than that of the flat shape, indicating that as the stress increases to the critical point, the rectangular membrane will experience a buckling transition, at which membrane starts to bend with an increasing bending energy $E_{\text{b}}$ at the cost of reduced boundary energy $E_{\text{l}}$. 

\begin{figure}[h]
\centering
\includegraphics[scale=0.26]{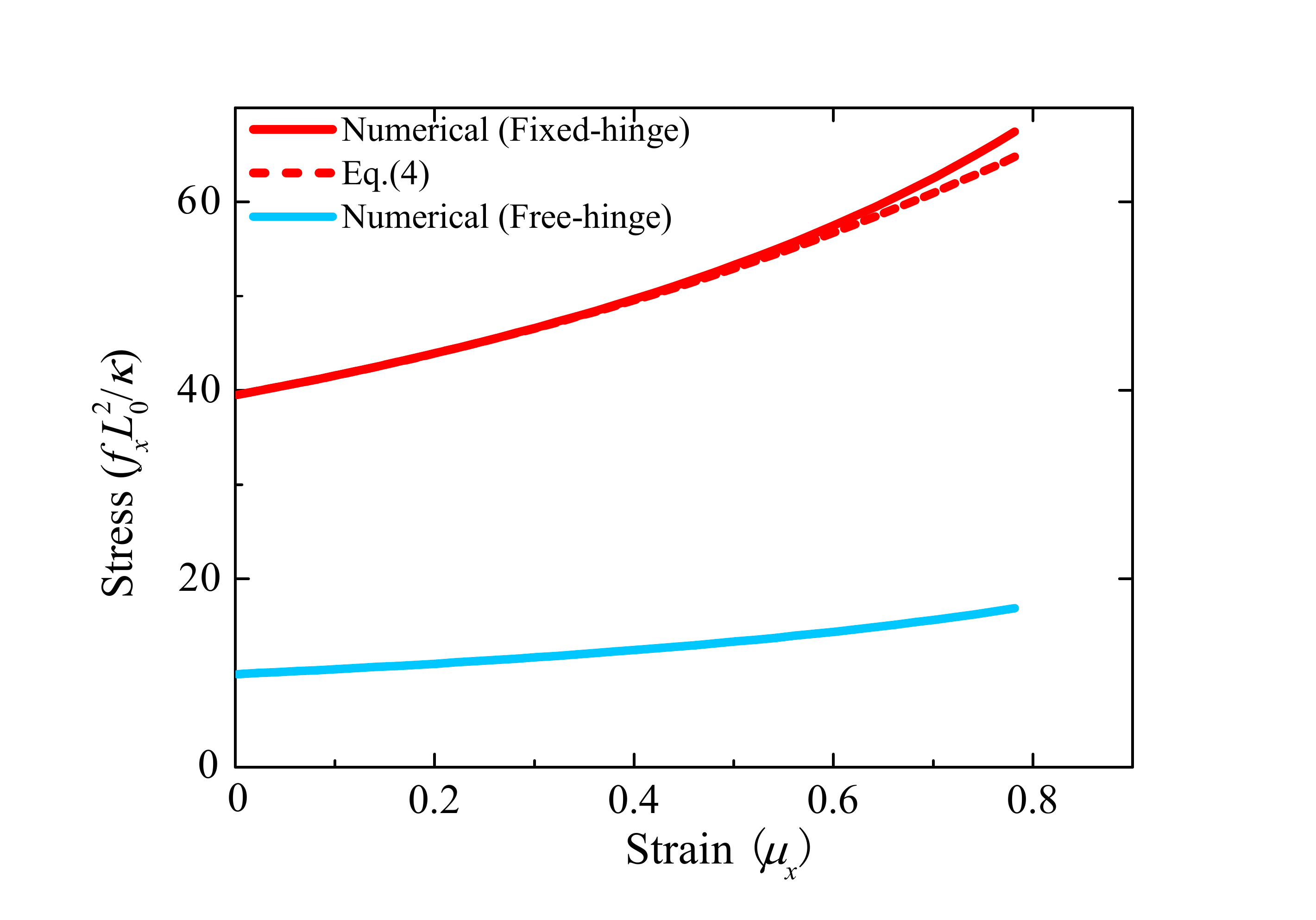}
\caption{Stress-strain relationship of the rectangular membrane. The dashed red curve represents the analytical result Eq.~(\ref{eq:buckle}) reported in Ref.~\cite{hu2013determining}. }
\label{fig4}
\end{figure}

Near the critical stress, the buckled membrane remains relatively flat. As the stress builds up, the membrane becomes more and more bent (top panels in Fig.~\ref{fig3}). In order to describe the bending degree, we define the strain $\mu_{x}=(L_{0}-L_{x})/{L_{0}}$ and plot the stress-strain relationship in Fig.~\ref{fig4}. Buckling is featured in the nonzero value of of the stress $f_x$ at $\mu_x = 0$. We find that the stress required to bend the membrane to the same strain for the fixed-hinge is 4-fold of that for the free-hinge. Furthermore, we compare our numerical results with the analytical results for the fixed-hinge BC derived in Ref.~\cite{hu2013determining},
\begin{equation}
\label{eq:buckle}
    f_x = \kappa \left(\frac{2\pi}{L_0}\right)^2\left[1+\frac{1}{2}\mu_x + \frac{9}{32}\mu_x^2+\frac{21}{128}\mu_x^3 + O(\mu_x^4)\right],
\end{equation}
and find a good agreement between them (the red solid line and the red dashed line in Fig.~\ref{fig4}).

\subsection{The buckling of a circular membrane}
In this section, we study the buckling process of a circular membrane under an isotropic and compressible radial stress $f_r$ with a unit of force per unit length. The membrane shape is assumed to remain axisymmetric upon buckling and thus can be depicted using its meridian profile. The effect of the Gaussian modulus is neglected, i.e., $\bar{\kappa} = 0$ for the moment and will be elaborated in the next section. It is found that the buckling behavior are qualitatively different for the two types of BCs. For the sake of comparison, we introduce the area strain $\mu_{r}=(A_{0}-A_{b})/{A_{0}}$ to reflect the buckling degree of the circular membrane, with $A_{0} = \pi R_0^2$ the total surface area and $A_{b} = \pi R_b^2$ the base area when the membrane buckles. 

\begin{figure}[!ht]
\centering
      \includegraphics[scale=0.45]{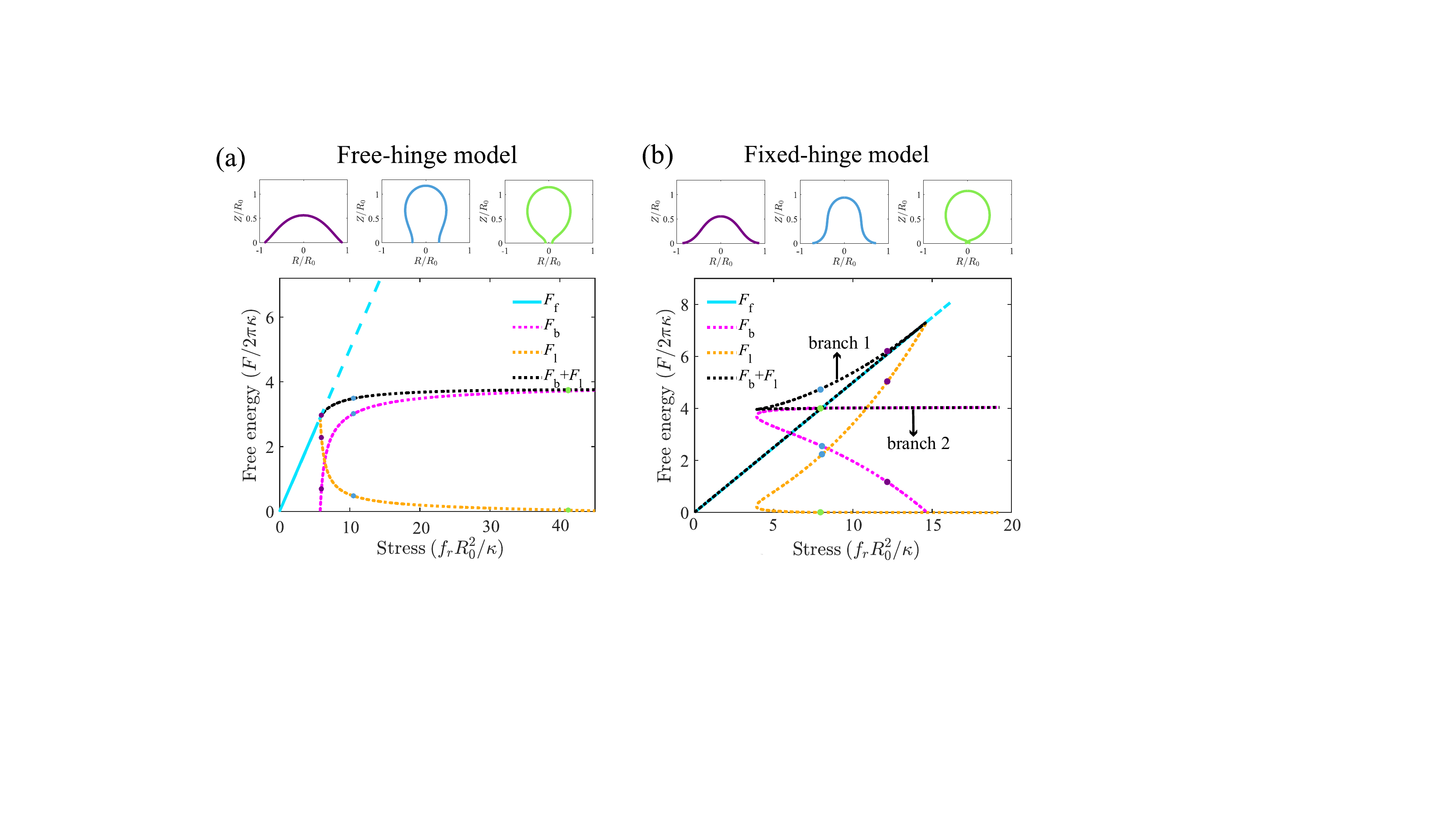}
\caption{Dependence of the total energy of the circular membrane on the stress based on (a) free-hinge BC and (b) fixed-hinge BC, respectively.}
\label{fig5}
\end{figure}

\begin{figure}[!ht]
\centering
      \includegraphics[scale=0.26]{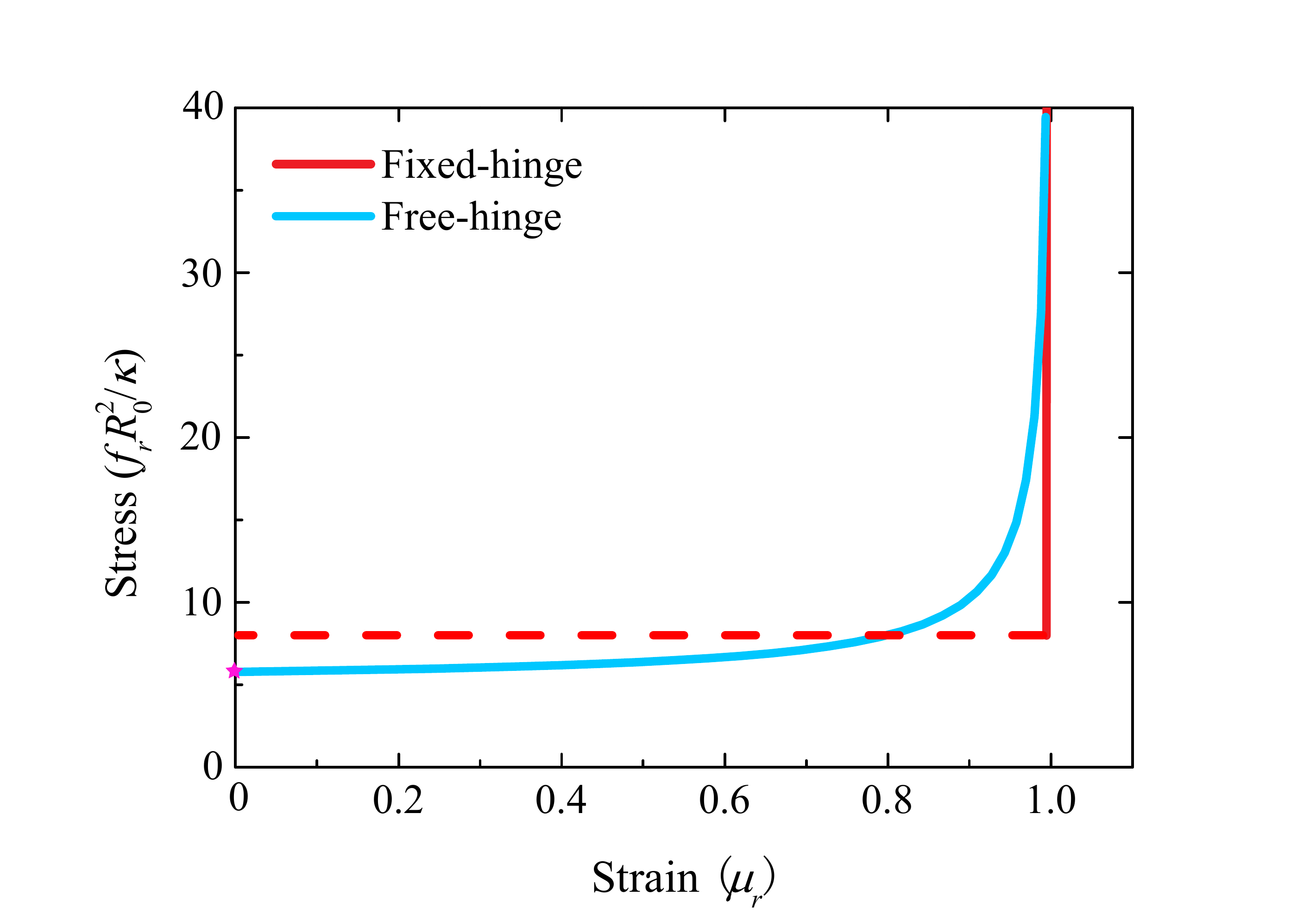}
\caption{Stress-strain relationship of the buckled circular membrane. The pentagon indicates the analytical result for the critical buckling stress given by Eq.~(\ref{eq:critstress}).}
\label{fig6}
\end{figure}

For the free-hinge BC, the buckling process is quite similar to that of the rectangular membrane. Upon increase of the stress $f_r$, a new branch of buckled solutions emerge in addition to the flat membrane (black dotted line and cyan line in Fig.~\ref{fig5}(a)). The total energy of the buckled membrane, consisting of the bending energy $F_{\text{b}}$ and the boundary energy $F_{\text{l}}$, is lower than that of the flat membrane $F_{\text{f}}$ , indicating the occurrence of a buckling transition. The membrane remains almost flat near the transition point, and after that the buckling degree continuously increases with the stress $f_r$ (top panels in Fig.~\ref{fig5}(a)), which is manifested as a continuous stress-strain relationship in the red curve of Fig.~\ref{fig6}. We analytically derive the critical stress 
\begin{equation}
\label{eq:critstress}
    f_r^{\mathrm{crit}}=\kappa \frac{\left[x_1^{(0)}\right]^2}{R_0^2},
\end{equation}
 a result in a good agreement with the numerical solution (the magenta pentagon in Fig.~\ref{fig6}).

For the fixed-hinge BC, the energy profiles become complicated and two buckled branches are found for a single stress $f_r$. On one of the buckled branches, the bending degree is decreased with the increasing stress $f_r$ (the purple and the blue shapes of the top panels in Fig.~\ref{fig5}(b)). Hereafter, we will refer to this branch as branch 1. While on the other branch, the base of the membrane is nearly closed (the green shape of the top panels in Fig.~\ref{fig5}(b)). Hereafter, we will refer to this branch as branch 2. The total energy, consisting of the bending energy $F_{\text{b}}$ and the boundary energy $F_{\text{l}}$, is lower in branch 2 than in branch 1. The energy of the flat membrane intersects with branch 2 at a point such that the membrane is nearly closed. All these results suggest that for the fixed-hinge BC, there exists a first-order transition at the critical stress, beyond which a sudden and sharp membrane buckling occurs. In the stress-strain relationship, it is reflected in the sudden jump of the strain from zero to almost 1 when the stress goes beyond the critical point (the blue curve in Fig.~\ref{fig6}).           

\begin{figure}[!ht]
\centering
      \includegraphics[scale=0.25]{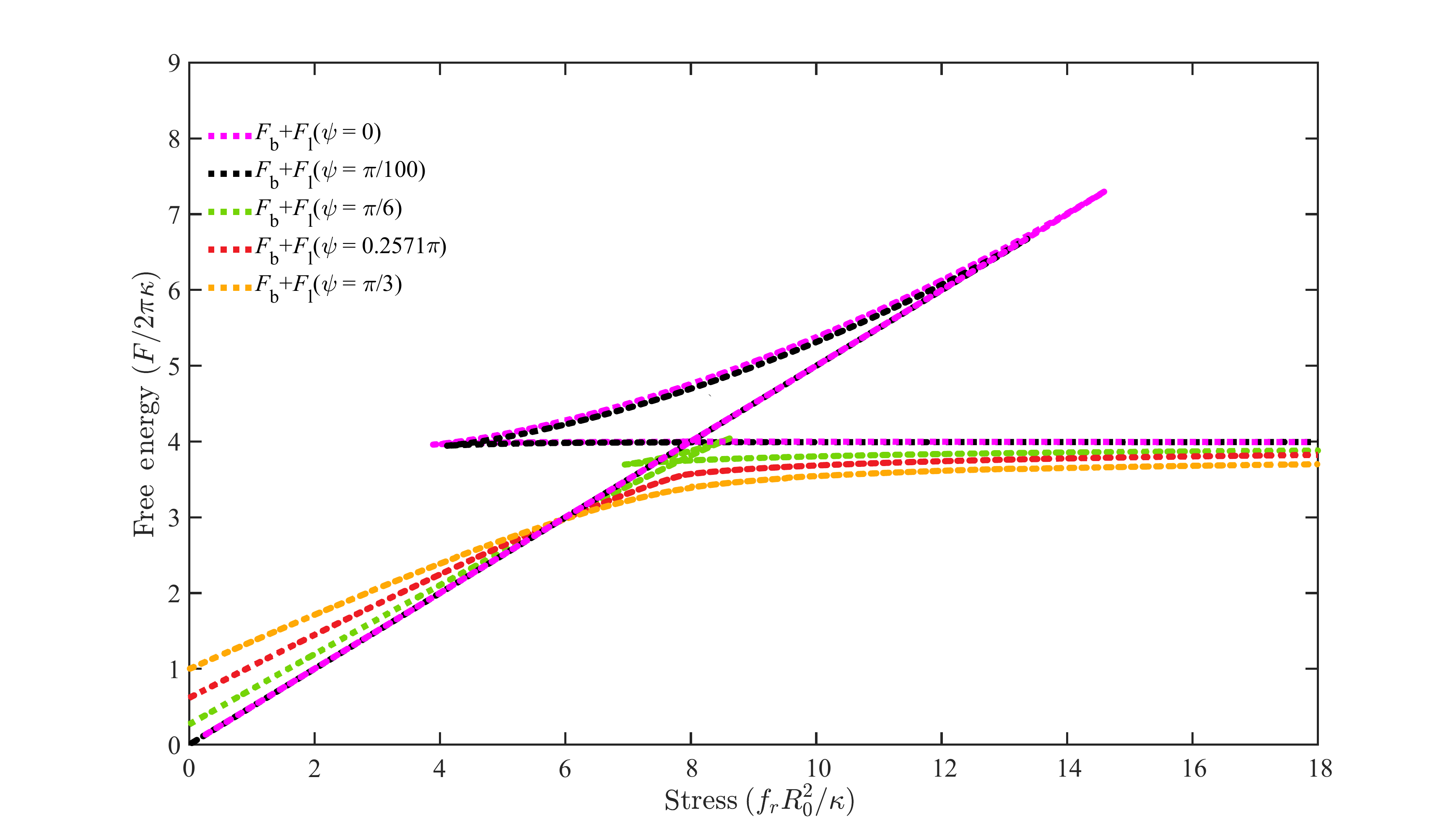}
\caption{Effect of the stress on the total energy of the circular membrane under the fixed-hinge BC for different hinged-angles.}
\label{fig7}
\end{figure}

To further understand the origin of the first-order transition under the fixed-hinge BC, we calculate the buckling process of a circular membrane under different hinged-angles. Note that for a nonzero hinged-angle, the membrane is already bent even at zero stress $f_r$. Upon increasing the stress $f_r$, if the hinged-angle is large, the total energy of the buckled membrane continuously increases with $f_r$ (the orange line in Fig.~\ref{fig7}). However, for small hinged angles, a Gibbs triangle appears in the energy profile (the magenta, black and green lines in Fig.~\ref{fig7}), which is the characteristic of a first-order transition. A further calculation tells us that the critical angle distinguishing between the first-order and the second order transitions is around $0.2571\pi$, as indicated by the red dotted line in Fig.~\ref{fig7}.

\subsection{Determination of the Gaussian curvature modulus via circular membrane buckling with a free-hinge BC}

In this section, we study the effect of the Gaussian curvature modulus $\bar{\kappa}$ on the buckling process of a circular membrane under the free-hinge BC condition Eq.~(\ref{eq:freehinge}), in which $\bar{\kappa}$ explicitly appears. By virtue of the fact that the Gaussian curvature modulus $\bar{\kappa}$ only contributes to the boundary bending moment, it makes no difference on the membrane shape equations, and therefore has no impact on the buckling process under the fixed-hinge BC in which $\psi = 0$ has no dependence on $\bar{\kappa}$. 

\begin{figure}[!ht]
\centering
      \includegraphics[scale=0.35]{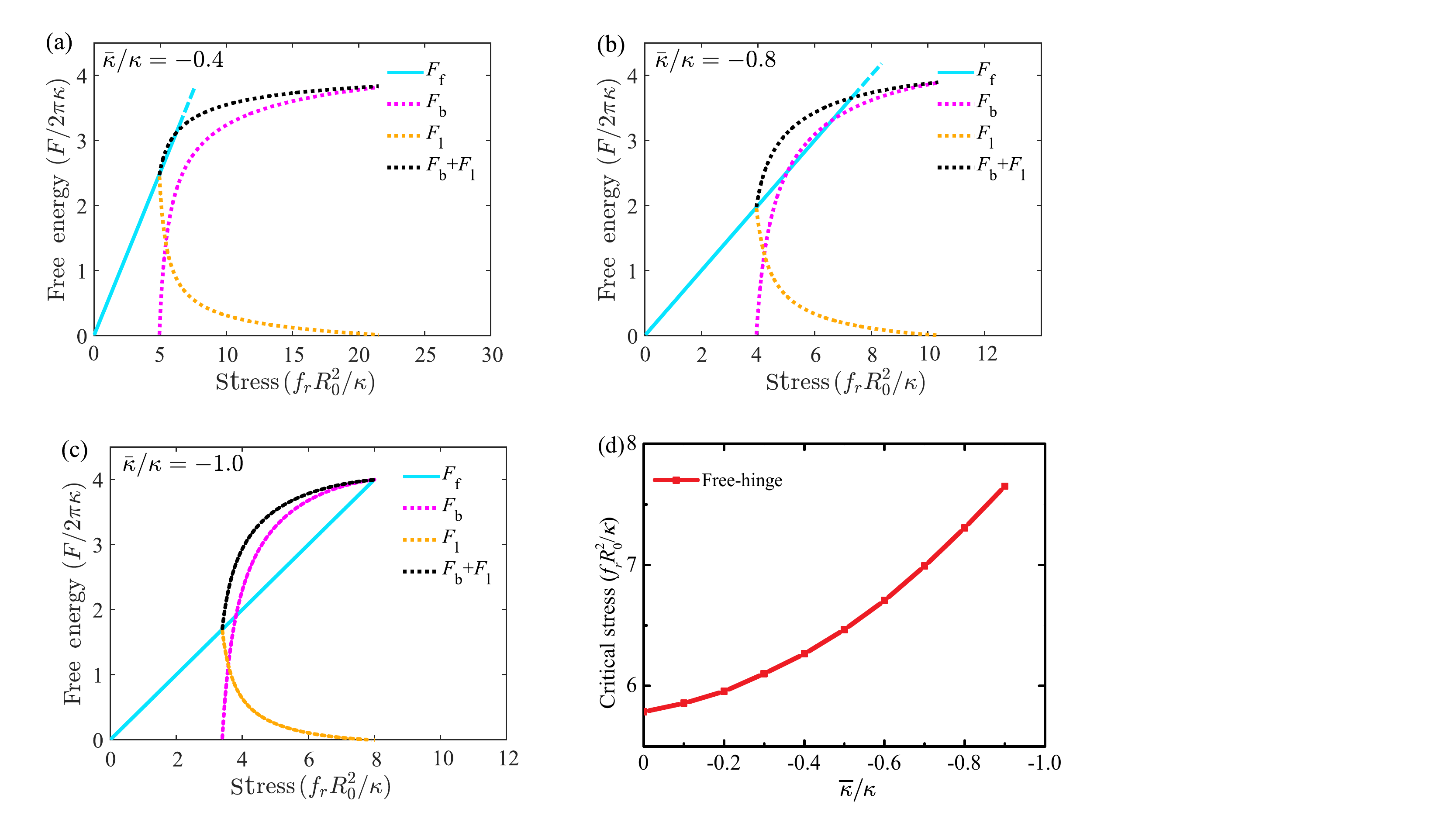}
\caption{Dependence of the free energy of the circular membrane on the stress under the free-hinge condition with three different Gaussian curvatures (a) $\bar{\kappa}/\kappa=-0.4$, (b) $\bar{\kappa}/\kappa=-0.8$ and (c) $\bar{\kappa}/\kappa=-1.0$. (d) The critical stress $f_{r}$ as a function of Gaussian curvature $\bar{\kappa}/\kappa$ under the free-hinge condition.}
\label{fig8}
\end{figure}

In Fig.~\ref{fig8}, we show the total energy $F = F_{\text{b}} + F_{\text{l}}$ of a buckled membrane as a function of the stress $f_r$ with different Gaussian curvature moduli under the free-hinge BC. It is found that the energy of the flat membrane intersects with the energy of the buckled membrane at two stresses (cyan lines in Fig.~\ref{fig8}~(a)-(c)). After the first stress, the energy of the flat membrane remains lower than that of the buckled one until the second stress, indicating that the critical buckling actually occurs at the second one. It is found this critical stress increases with the absolute value of the Gaussian curvature modulus $\left| \bar{\kappa}/{\kappa}\right|$, as shown in Fig.~\ref{fig8}(d).

\begin{figure}[!ht]
\centering
      \includegraphics[scale=0.3]{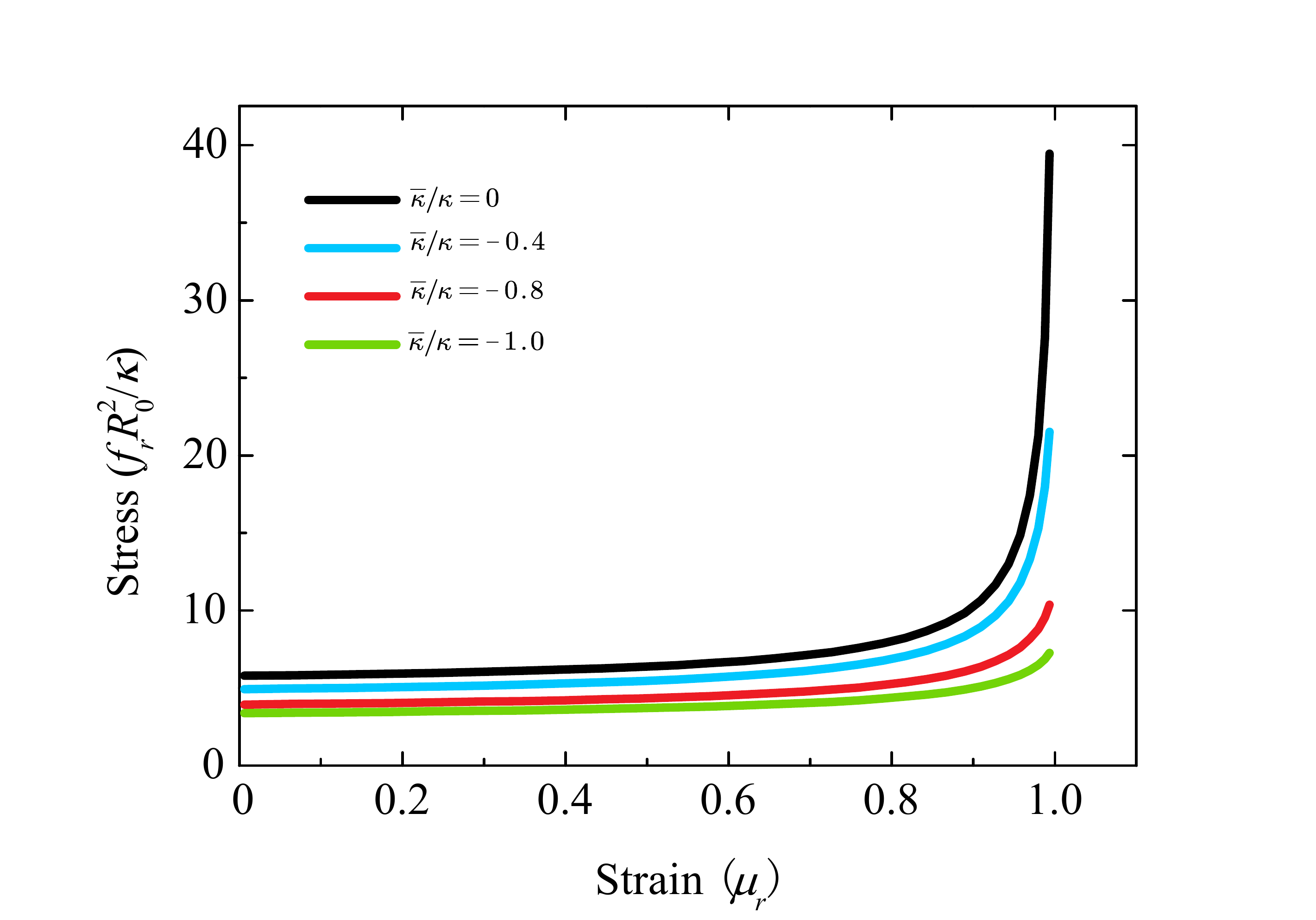}
\caption{The stress-strain relationship of the circular membrane with different Gaussian curvatures.}
\label{fig9}
\end{figure}

In principle, the Gaussian bending modulus can be determined from Fig.~\ref{fig8}(d) if we can measure the critical stress to buckle a circular membrane under the free-hinge BC from MD simulations. However, the precision of the measurement will be limited by the single variable measurement. To overcome this issue, we see that the stress-strain relationship for different Gaussian curvature moduli are quite different (Fig.~\ref{fig9}). Fitting the stress-strain relationship of the MD simulation with the numerical results at multiple stresses therefore provides a more robust way to estimate the Gaussian curvature modulus $\bar{\kappa}$. Compared with measuring the fluctuation spectrum at the edge, obtaining the stress-strain relationship is more straightforward. We speculate the method is also robust against the coarse graining level of the lipid models and the treatment of the solvent based on its performance in the measurement of the mean curvature modulus $\kappa$~\cite{hu2013determining}. Test of the method with MD simulations will be one of our future works. 

\section{conclusion}
In summary, we investigate the buckling of a rectangular membrane and a circular membrane under two BCs. It is found that for an initially flat rectangular membrane, when the stress is increasingly loaded, buckling occurs continuously under both free-hinge and fixed-hinge BCs. But for the initially flat circular membrane, the buckling behavior depends on its boundary condition. For the free-hinge BC, with the increase of stress, buckling takes place continuously, a process the same as that of the rectangular membrane. But for the fixed-hinge BC, there exists a first-order buckling transition if the hingedangle is small. Finally, we find an effective method to determine the Gaussian curvature modulus by using the stress-strain relationship of the circular membrane under the free-hinge BC. 
\begin{acknowledgments}
We acknowledge financial support from National Natural Science Foundation of China under Grants No. 12147142, No. 11974292, No. 12174323, and No. 12004317, Fundamental Research Funds for Central Universities of China under Grant No. 20720200072 (RM), and 111 project No. B16029.
\end{acknowledgments}

\appendix

\section{Theoretical details for rectangular membranes}
For the rectangular membrane, one can obtain its principal curvature $c_1 = \dot{\psi}$. The elastic energy \emph{E} in Eq.~(\ref{recttotalenergy}) then can be written as $ E=\kappa L_{y}\int_0^{L_0/2}{\mathcal{L}\text{d}s} $, with the Lagrangian function 
\begin{equation}
\mathcal{L}=\frac{1}{2} \dot{\psi} ^2 +\bar{f_{x}}\cos \psi +\gamma \left( \dot{X}-\cos \psi \right) +\eta \left( \dot{Z}+\sin \psi \right),
\end{equation}
where $\bar{f_{x}}=f_{x} /\kappa $, $\dot{\psi}$ and $\dot{X}$ denote their derivatives with respect to the arclength $s$, and $\gamma$ and $\eta$ are the Lagrangian multipliers that enforce the geometric relations 
\begin{equation}
\label{eq:geometry}
    \dot{X}=\cos \psi, \quad\quad \dot{Z}=-\sin \psi.
\end{equation}
A variation of the functional \emph{E} gives
\begin{equation}
\begin{aligned}
\delta E&=\int_0^{L_0/2}{\text{d}s}\left\{ \left[ \frac{\partial \mathcal{L}}{\partial \psi}-\frac{\text{d}}{\text{d}s}\frac{\partial \mathcal{L}}{\partial \dot{\psi}} \right] \delta \psi +\left[ \frac{\partial \mathcal{L}}{\partial X}-\frac{\text{d}}{\text{d}s}\frac{\partial \mathcal{L}}{\partial \dot{X}} \right] \delta X \right.\\
&\left.+\left[ \frac{\partial \mathcal{L}}{\partial Z}-\frac{\text{d}}{\text{d}s}\frac{\partial \mathcal{L}}{\partial \dot{Z}} \right] \delta Z+\frac{\partial \mathcal{L}}{\partial \gamma}\delta \gamma +\frac{\partial \mathcal{L}}{\partial \eta}\delta \eta \right\}\\ 
&-H\delta s\mid_{0}^{L_0/2}+\frac{\partial \mathcal{L}}{\partial \dot{\psi}}\delta \psi \mid_{0}^{L_0/2}+\frac{\partial \mathcal{L}}{\partial \dot{X}}\delta X\mid_{0}^{L_0/2}+\frac{\partial \mathcal{L}}{\partial \dot{Z}}\delta Z\mid_{0}^{L_0/2}, 
\label{rectvariation}
\end{aligned}
\end{equation}
where $ H\equiv -\mathcal{L}+\dot{\psi}\partial \mathcal{L}/\partial \dot{\psi}+\dot{X}\partial \mathcal{L}/\partial \dot{X}+\dot{Z}\partial \mathcal{L}/\partial \dot{Z} $ is the Hamiltonian function given by 
\begin{equation}
H = \frac{1}{2}\dot{\psi}^2 -\bar{f_{x}} \cos \psi+\gamma \cos \psi -\eta \sin \psi.
\end{equation}
Having the bulk terms of Eq.~(\ref{rectvariation}) vanish leads to the shape equations
\begin{equation}
\label{eq:ddpsi}
\ddot{\psi}=\dot{\psi}+\gamma\sin \psi +\eta\cos \psi -\bar{ f_{x}}\sin \psi
\end{equation}
\begin{equation}
\label{eq:dgamma}
\dot{\gamma}=0
\end{equation}
\begin{equation}
\label{eq:deta}
\dot{\eta}=0.
\end{equation}
Having the boundary terms in Eq.~(\ref{rectvariation}) vanish, one can obtain the BCs. In particular, at the membrane tip $\emph{s}=0$, we have 4 BCs: $\psi(0)=0$, $X(0)=0$, $\gamma(0)=\bar{f_{x}}-[\dot{\psi}(0)]^2/2$ and $\eta(0) = 0$. At the membrane base $s=L_0/2$, we have 3 BCs: $Z(L_0/2)=0$, $X(L_0/2)=L_x/2$, and $\psi(L_0/2)=0$ for the fixed-hinge BC or $\dot{\psi}(L_0/2)=0$ for the free-hinge BC. 

In summary, Eqs.~(\ref{eq:geometry}), (\ref{eq:ddpsi}), (\ref{eq:dgamma}), and (\ref{eq:deta}) constitute the full set of shape equations for rectangular membranes. They can be converted to 6 first order ordinary differential equations. Together with the unknown Lagrangian multiplier $\bar{f}_x$ and the 7 BCs, we solve the problem with the bvp5c solver in MATLAB that is designed for solving boundary value problems of ordinary differential equations. 

\section{Theoretical details for circular membranes}
For the circular membrane, the meridian coordinates $ R\left( s \right) $ and $Z\left( s \right) $ satisfy the geometric relations via
\begin{equation}
\label{eq:geocirc}
\dot{R}=\cos \psi, \quad\quad  \dot{Z}=-\sin \psi.
\end{equation}
The elastic energy $\emph{F}$ in Eq.~(\ref{cirtotalenergy}) can be expressed as $ F=2\pi \kappa \int_0^S{\mathcal{L}\text{d}s} $, with the Lagrangian function
\begin{equation}
\begin{aligned}
\mathcal{L}&=\frac{R}{2}\left( \dot{\psi}+\frac{\sin \psi}{R} \right) ^2+\bar{\kappa}\dot{\psi}(\sin\psi)/\kappa+\bar{\sigma}R+{\bar{f_{r}} R}\cos \psi\\
&+\gamma \left( \dot{R}-\cos \psi \right)+\eta \left( \dot{Z}+\sin \psi \right),
\end{aligned}
\end{equation}
where $ \bar{\sigma}=\sigma /\kappa $, $ \bar{f_{r}}=f_{r} /\kappa $, $\gamma(s)$ and $\eta(s)$ are Lagrangian multipliers to enforce the geometric relations in Eq.~(\ref{eq:geocirc}). The total arclength $S$ is an unknown parameter to be solved with shape equations.
A variation of the functional \emph{F} reads
\begin{equation}
\begin{aligned}
\delta F&=\int_0^S{\text{d}s}\left\{ \left[ \frac{\partial \mathcal{L}}{\partial \psi}-\frac{\text{d}}{\text{d}s}\frac{\partial \mathcal{L}}{\partial \dot{\psi}} \right] \delta \psi +\left[ \frac{\partial \mathcal{L}}{\partial R}-\frac{\text{d}}{\text{d}s}\frac{\partial \mathcal{L}}{\partial \dot{R}} \right] \delta R \right.\\
&\left.+\left[ \frac{\partial \mathcal{L}}{\partial Z}-\frac{\text{d}}{\text{d}s}\frac{\partial \mathcal{L}}{\partial \dot{Z}} \right] \delta Z+\frac{\partial \mathcal{L}}{\partial \gamma}\delta \gamma +\frac{\partial \mathcal{L}}{\partial \eta}\delta \eta \right\}\\
&-H\delta s\mid_{0}^{S}+\frac{\partial \mathcal{L}}{\partial \dot{\psi}}\delta \psi \mid_{0}^{S}+\frac{\partial \mathcal{L}}{\partial \dot{R}}\delta R\mid_{0}^{S}+\frac{\partial \mathcal{L}}{\partial \dot{Z}}\delta Z\mid_{0}^{S},
\label{cirvariation}
\end{aligned}
\end{equation}
where the Hamiltonian function $ H\equiv -\mathcal{L}+\dot{\psi}\partial \mathcal{L}/\partial \dot{\psi}+\dot{R}\partial \mathcal{L}/\partial \dot{R}+\dot{Z}\partial \mathcal{L}/\partial \dot{Z} $ can be expressed as
\begin{equation}
\begin{aligned}
H&= \frac{R}{2}\left[ \dot{\psi}^2-\left( \frac{\sin \psi}{R} \right) ^2 \right] -\bar{\sigma}R-\bar{f_{r}}R\cos \psi\\
&+\gamma \cos \psi -\eta \sin \psi =0.
\end{aligned}
\end{equation}
If we have the bulk terms of Eq.~(\ref{cirvariation}) vanish, we obtain the following shape equations
\begin{equation}
\label{eq:circddpsi}
\ddot{\psi}=\frac{\sin \psi \cos \psi}{R^2}-\frac{\dot{\psi}}{R}\cos \psi+\frac{\gamma}{R}\sin \psi +\frac{\eta}{R}\cos \psi -\bar{f_{r}}\sin \psi,
\end{equation}
\begin{equation}
\label{eq:circdgamma}
\dot{\gamma}=\frac{1}{2} \dot{\psi} ^2-\frac{\sin ^2\psi}{2R^2}+\bar{\sigma}+\bar{f_{r}}\cos \psi,
\end{equation}
\begin{equation}
\label{eq:circdeta}
\dot{\eta}=0.
\end{equation}
The BCs can be obtained by setting the boundary terms in Eq.~(\ref{cirvariation}) to be zero. In particular, at the membrane tip \emph{s}=0, we have 4 BCs: $\psi(0)=0$, $R(0)=0$, $\gamma(0)=0$, and $\eta(0)=0$. At the membrane base $s=S$,  we have 3 BCs: $Z(S)=0$, $R(S)=R_b$ , $\psi(S)=0$ for the fixed-hinge BC or $\kappa(\dot{\psi}+\sin\psi/R)+\bar{\kappa}\sin\psi/R=0$ for the free-higne BC. In addition, we need to impose the incompressibility condition
\begin{equation}
\label{eq:area}
    2\pi\int_0^Srds = \pi R_0^2. 
\end{equation}

In summary, Eqs.~(\ref{eq:geocirc}), (\ref{eq:circddpsi}), (\ref{eq:circdgamma}), and (\ref{eq:circdeta}) make up the full set of shape equations for circular membranes. They can be converted to 6 first order ordinary differential equations. Together with the 2 unknown parameter $\bar{f}_r$ and $S$, as well as the 7 BCs and the incomprehensibility constraint (\ref{eq:area}), we can solve the problem with the MATLAB solver bvp5c.

\section{Analytical results for the critical buckling stress of a circular membrane under the free-hinge BC condition}
For an almost flat membrane, the angle $\psi \ll 1 $. Under this approximation, we can get the linearized shape equation
\begin{equation}
\label{eq:linear}
R^2\psi ''+R\psi '-\left( R^2\bar{\sigma}+1 \right) \psi =0,
\end{equation}
where we have converted the function $\psi(s)$ to $\psi(R)$ with the prime denoting the derivative with respect to $R$. The equation has a physically meaningful solution
\begin{equation}
\label{eq:bessel}
\psi \left( R \right) =C_1J_1\left( R\sqrt{-\bar{\sigma}} \right),
\end{equation}
 only if $\bar{\sigma}<0$. Here $C_1$ is an arbitrary constant and $J_1(x)$ denotes the first kind of Bessel function. The free-hinge BC requires that the following equation
\begin{equation}
\label{eq:linearbc}
    \psi'+\frac{\psi}{R}=0
\end{equation}
holds at $R=R_b$. Substituting Eq.~(\ref{eq:bessel}) into (\ref{eq:linearbc}), we obtain
\begin{equation}
J_0\left(R_b\sqrt{-\bar{\sigma}} \right)=0.
\end{equation}
In order to get the first buckling mode, we let $R_b\sqrt{-\bar{\sigma}}=x_{1}^{\left( 0 \right)}$, the first zero value of the Bessel function $J_0(x)$. The resulting membrane tension reads
\begin{equation}
\label{eq:sigma}
\bar{\sigma}= -\frac{1}{R_{b}^{2}}\left[ x_{1}^{\left( 0 \right)} \right] ^2.
\end{equation}
 The critical stress at which buckling occurs is essentially the negative tension in Eq.~(\ref{eq:sigma}) and have the base radius $R_b = R_0$. In this way, we obtain Eq.~(\ref{eq:critstress}).
 

\nocite{*}

\bibliography{apssamp}

\end{document}